\documentclass[twocolumn]{aastex631}
\received{December 22, 2022}
\revised{January 9, 2023}
\accepted{January 9, 2023}

\begin{document}

\title{Astrophysical properties of 600 bonafide single stars in the Hyades open cluster}

\author[0000-0003-1939-6351]{Wolfgang Brandner}
\affiliation{Max-Planck-Institut f\"ur Astronomie\\ K\"onigstuhl 17\\ 69117 Heidelberg, Germany}

\author{Per Calissendorff}
\affiliation{Department of Astronomy\\ University of Michigan\\Ann Arbor, MI 4810, USA}

\author{Taisiya Kopytova}
\affiliation{Ural Federal University\\ Yekaterinburg\\ 620002, Russia}
\affiliation{Max-Planck-Institut f\"ur Astronomie\\ K\"onigstuhl 17\\ 69117 Heidelberg, Germany}

\correspondingauthor{Wolfgang Brandner}
\email{brandner@mpia.de}



\begin{abstract}
The determination of the astrophysical properties of stars remains challenging, and frequently relies on the application of stellar models. Stellar sequences in nearby open clusters provide some of the best means to test and calibrate stellar evolutionary models and isochrones, and to use these models to assign astrophysical properties consistently to a large sample of stars.
We aim at updating the single star sequence of members of the Hyades cluster, identifying the best-fitting isochrones, and determining the astrophysical properties of the stars.
The Gaia Catalogue of Nearby Stars provides a comprehensive sample of high-probability members of the Hyades cluster. We apply a multi-step method to flag photometric outliers, and to identify bonafide single stars and likely binary and multiple systems. 
 The single stars define a tight sequence, which in the mass range 0.12 to 2.2\,M$_\odot$ is well-fitted by PARSEC isochrones for a supersolar metallicity of [M/H] = $+0.18\pm 0.03$ and an age of $775\pm25$\,Myr. The isochrones enable us to assign mass, effective temperature, luminosity, and surface gravity to each of the 600 bonafide single main-sequence stars. 
The observed sequence validates the PARSEC isochrones. The derived stellar properties can serve as benchmarks for atmospheric and evolutionary models, and for all-sky catalogs of stellar astrophysical properties. The stellar properties are also relevant for studies of exoplanet properties among Hyades exoplanet hosts.
\end{abstract}



\section{Introduction} \label{sec:intro}

Studies of proto-cluster environments and Galactic starburst clusters at ages well below 10\,Myr suggest that clustered star formation occurs quasi instantaneous, with a maximum age spread among cluster members 
of the order of 0.1 to 0.4\,Myr \citep{Kudryavtseva2012,Parmentier2014,Williams2022}. Galactic open clusters with typical ages of a few to several 100\,Myr also exhibit small spreads in age and metallicity among their members, suggesting close-to-coeval formation out of the homogeneously mixed material of the parental giant molecular cloud \citep{Bovy2016,Magrini2017,Jeffries2017,Krumholz2020}.

Star clusters are thus suited particularly well for studying stellar astrophysical properties as a function of stellar mass, or its proxies stellar effective temperature and luminosity. Due of its proximity and richness in stellar members, the nearby Hyades open cluster is frequently used for testing and calibrating stellar evolutionary models \citep{Castellani2001,Kopytova2016,Jeffery2022}.

Based on Gaia Early Data Release 3 (EDR3) astrometry and radial velocities, \cite{GAIA_Smart2021A} identify more than 3000 potential members of the Hyades cluster within 100\,pc of the Sun. After application of a local density filter, they  reduce this to a sample of 920 candidate members. In \cite{Brandner2023} we use this sample as a basis for benchmarking MESA evolutionary models \citep{Choi2016,Dotter2016}, and find a significant discrepancy between the observational sequence and theoretical isochrones in the stellar mass range 0.25 to 0.85\,M$_\odot$. 

A possible fix to resolve this discrepancy is a modification of the temperature-Rosseland mean optical depth (T-$\tau$) relation in the stellar models, as introduced by \cite{Chen2014} for the PAdova and TRieste Stellar Evolution Code (PARSEC, \cite{Giradi2002,Marigo2008,Bressan2012}). 

In the present paper, we test PARSEC isochrones against the cleaned-up single star sequence of the Hyades open cluster, and use the family of best fitting isochrones to assign astrophysical properties to 600 candidate members of the Hyades. The structure of the paper is as follows: in section 2 we describe the processing of the Gaia data and the identification of the single stars. In section 3 we identify the best fitting PARSEC isochrones, and determine the stellar properties. 
In section 4 we close with a discussion and outlook.



\section{The single star sequence: cleaning the Gaia data set} \label{sec:sss}

While in general of exquisite and unprecedented quality, Gaia DR3 includes some problematic data. This in particular concerns bright sources saturating too many pixel on the Gaia detectors, and faint sources, which can be subject to photometric blends in particular in the spectrally dispersed BP and RP bands. The {\it Gaia Catalogue of Nearby Stars} (GCNS) also includes faint red stars with significant detections in the G and RP bands, and with BP magnitudes at or below the formal Gaia detection limit \citep{GAIA_Smart2021A}. 

\begin{figure*}[ht!]
\plotone{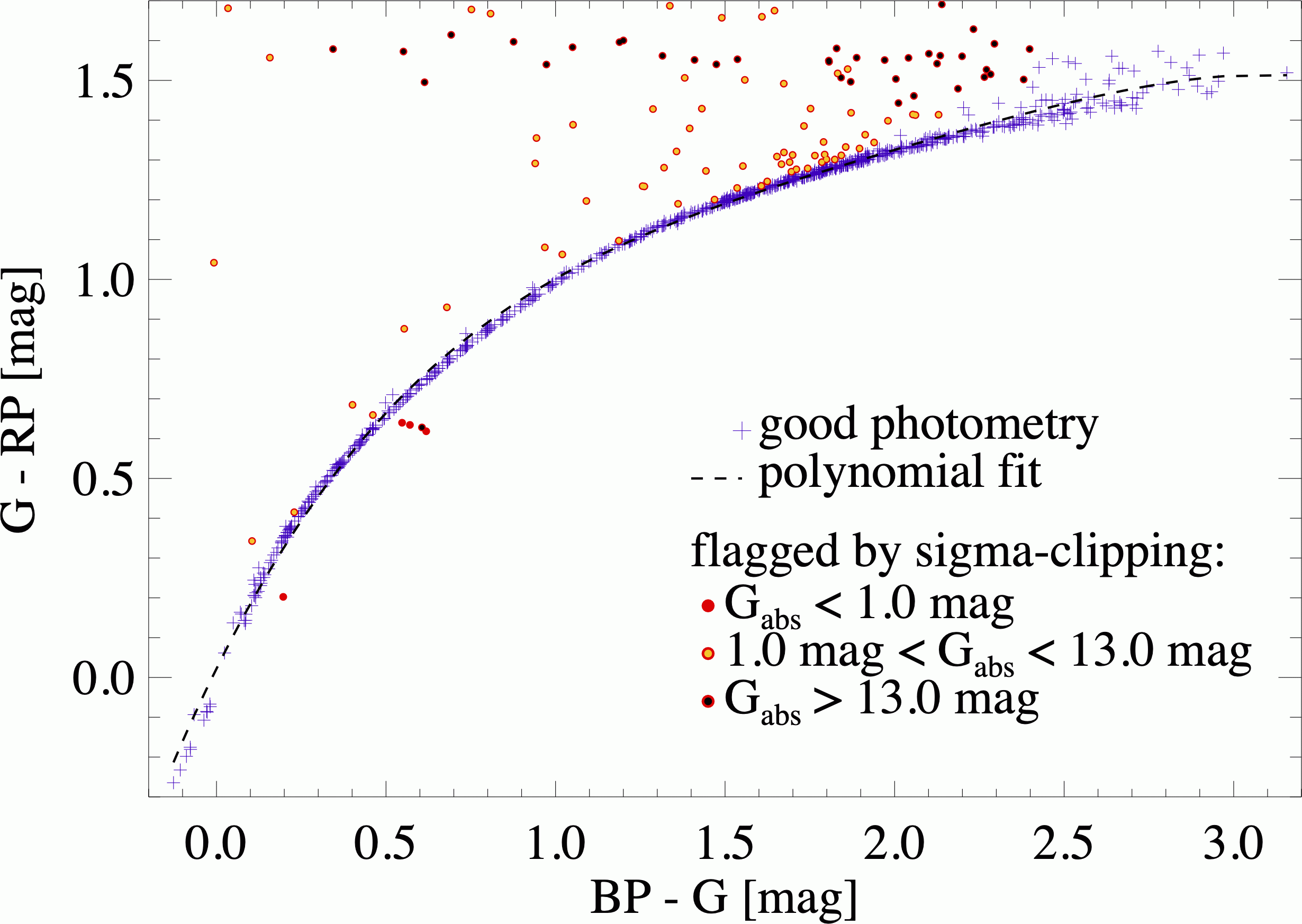}
\caption{Color-color diagram of the Hyades open cluster based on Gaia DR3 photometry. Blue plus signs mark sources with valid photometry in all three Gaia photometric bands. Red circles mark stars with potentially unreliable photometry in at least one of the photometric bands as identified by sigma-clipping.
\label{fig:HyaTCD}}
\end{figure*}

\begin{figure}[ht!]
\plotone{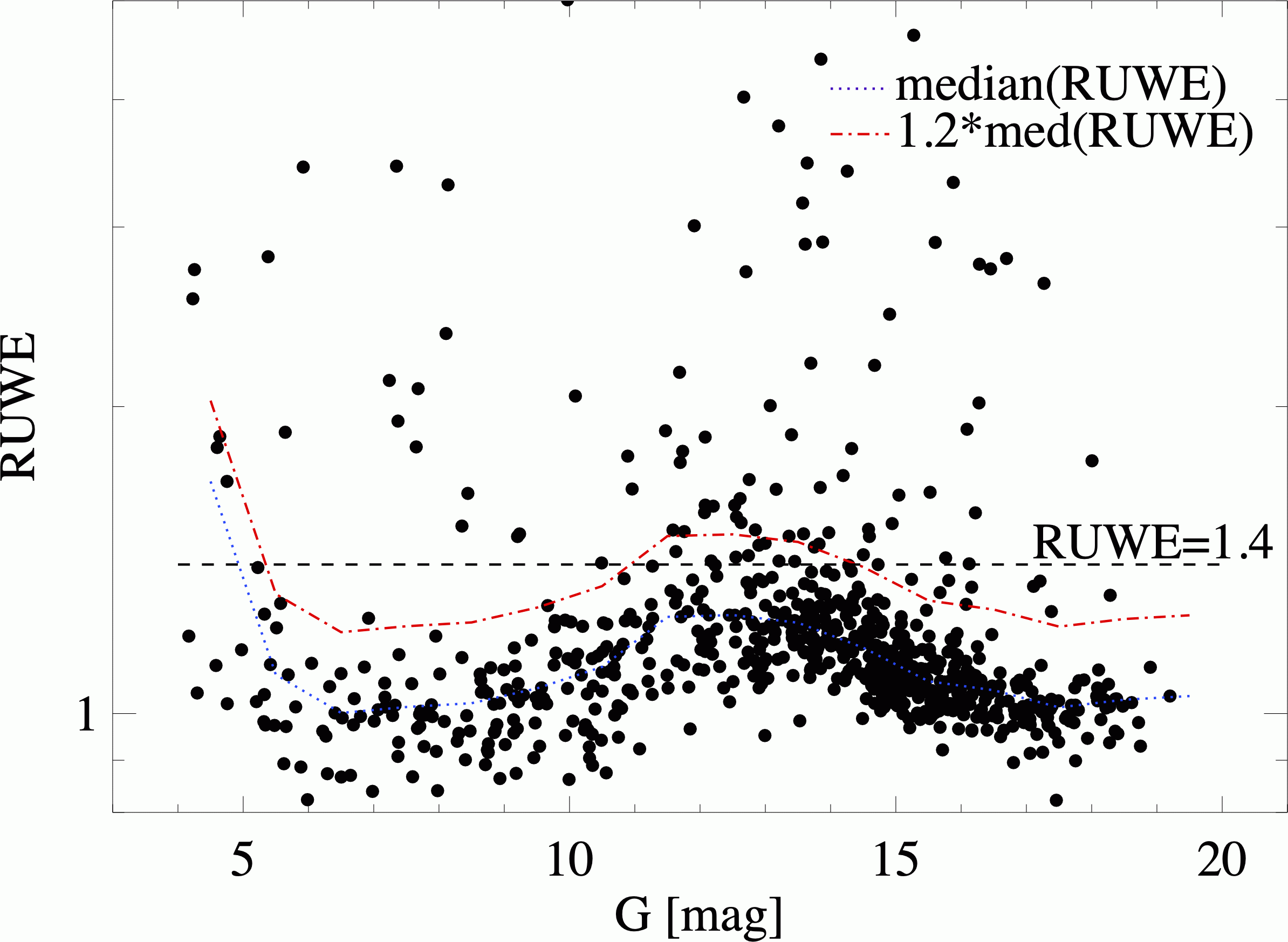}
\caption{RUWE as a function of G. The dashed line marks the canonical borderline between likely binary and multiple systems with RUWE$>$1.4, and bonafide single stars. For comparison the dotted blue and dash-dotted red lines mark the sliding window median, and 1.2 times sliding window median RUWE values.
\label{fig:HyaRUWE}}
\end{figure}

In \cite{Brandner2023} we defined a multi-step process to classify the 920 GCNS candidate members of the Hyades open cluster into four groups: bonafide single stars, probable binary and multiple systems, white dwarfs, and sources with problematic photometry in at least one of the Gaia photometric bands. In order to flag sources with potentially unreliable photometry, we iteratively fit a 4th order polynomial to the Gaia DR3\footnote{In \cite{Brandner2023} we used Gaia EDR3 G magnitudes, which had not been corrected for the processing error described in section 8.3 of \cite{Riello2021}. This error, which amounts to up to 0.0025\,mag, affected 100 of the 920 GCNS candidate members of the Hyades.} data in G-RP vs.\ BP-G color-color space, and then apply sigma-clipping at the 3$\sigma$ level (Figure \ref{fig:HyaTCD}), resulting in 783 sources with reliable photometry. The stars flagged with potentially unreliable photometry fall into three classes: i) four giant stars ($\delta$\,Tau, $\epsilon$\,Tau, $\gamma$\,Tau, and $\Theta^2$\,Tau, filled red circles) in the Hyades with G$_{\rm abs} < 1$\,mag, which are brighter than the nominal saturation limit of Gaia (E)DR3 (\cite{Riello2021}, Appendix C.1). ii) 38 stars with G$_{\rm abs} > 13$\,mag (red circles with yellow (light) inserts). These stars have $<$G$>_{\rm median} \approx$19.5\,mag and an apparent BP brightness below the nominal Gaia epoch detection threshold of BP=20.3\,mag, which results in an overestimate of the mean flux (\cite{Riello2021}, section 8.1). In the two color diagram, these stars occupy a band around G-RP$\approx$1.5 to 1.6\,mag, the only exception being Gaia EDR3 1988661087451017856, which has a photo-geometric distance of $\approx$96\,pc \citep{Bailer2021} and is classified as a candidate white dwarf by \cite{Gentile2021}. iii) 67 stars with 1\,mag $<$ G$_{\rm abs} < 13$\,mag, which have $<$G$>_{\rm median} \approx$13.9\,mag (red circles with black (dark) inserts). In the two color diagram these stars are all located to the left (bluer BP-G color) and above (redder G-RP color) of the typical sequence defined by the photospheres of Hyades members. We were not able to assign any plausible astrophysical reason for their peculiar Gaia colors.

Next we use the Re-normalized Unit Weight Error (RUWE, see \cite{gaia_edr3lite,Lindegren2021}) to classify stars with RUWE$<$1.4 as bonafide single stars. The sensitivity of the astrometric selection varies with projected separation, orbital period, and mass ratio of the binaries. Very close binaries, e.g., are in general not identified by the RUWE selection. In a color-absolute magnitude diagram, though, binaries composed of two main sequence stars stand out as being brighter sources located above and potentially redwards of the single star main sequence (see, e.g., \cite{Elson1998MNRAS}). Similary, binaries composed of a main-sequence star and a white dwarf (such as HZ 9, see below) would be located bluewards (and below) of the single star main sequence. In order to reject these sources from the sequence of bonafide single stars, we iteratively fitted an 8th order polynomial and applied sigma-clipping at the 3$\sigma$ level in a color-absolute magnitude diagram to all sources classified to have good photometry, and with RUWE$\le$ 1.4. After the convergence of the iterative process, the majority of the sources on the binary sequence are flagged as likely binaries. 

\cite{Golovin2022} discuss that the distinction between likely binary and multiple systems and bonafide single stars at RUWE = 1.4 irrespective of G band magnitude or BP - RP color might introduce a bias in terms of sample completeness. For the Hyades sample, though, the effect is small. Using, e.g., 1.2 times the sliding window median value of RUWE as function of G as the dividing line would add 3 stars with G$<$5\,mag and 8 stars with 10.5$<$G$<$14.5\,mag to -- and remove 10 stars with G$>$15\,mag from -- our sample of $\approx$600 bonafide single stars (Figure \ref{fig:HyaRUWE}).


\begin{figure*}[ht!]
\plotone{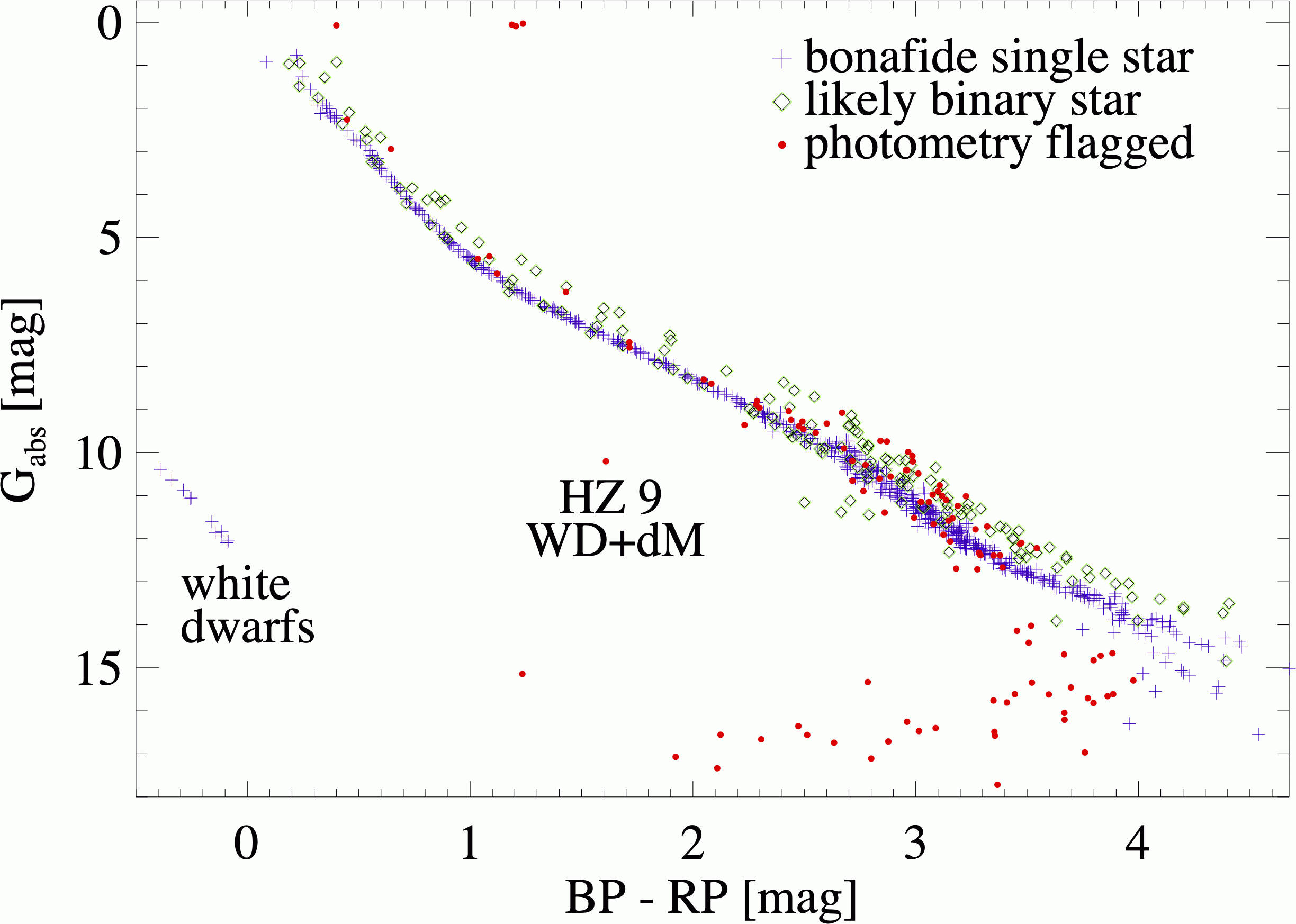}
\caption{Color-absolute magnitude diagram of candidate members of the Hyades open cluster based on GAIA DR3, showing the results of our classification into bonafide single main-sequence stars (and white dwarfs), likely binary and multiple systems, and stars with problematic photometry in at least one of the Gaia bands.
\label{fig:HyaCMDall}}
\end{figure*}

For the transformation to absolute G magnitudes, we use the {\it photgeo} distance estimates from \cite{Bailer2021}.
In Figure \ref{fig:HyaCMDall} we show the color-absolute magnitude diagram of the Hyades after the completion of the classification process. Of the sources with good photometry, we classify 601 sources as bonafide hydrogen burning single stars, 171 sources as likely binary stars, and 11 sources as single white dwarfs with BP - RP $<$ 0\,mag. The single stars form a tight and well-defined sequence with median uncertainties of $<\sigma_{\rm G_{abs}}>$ = 3.3\,mmag and $<\sigma_{\rm BP-RP}>$ = 4.1\,mmag, which presents a factor of $\approx$20 reduction of uncertainty in absolute magnitude and color compared to \cite{Kopytova2016}. In addition to the single white dwarfs, in Figure \ref{fig:HyaCMDall} we also label the post-common envelope white dwarf -- mid-M dwarf binary HZ 9 \citep{Rios2020}. This source is misclassified by our two color rejection due to its atypical composite spectral energy distribution, resulting in BP-G=0.68\,mag and G-RP=0.93\,mag. Nevertheless, HZ 9 would be excluded from the single star sample due to its binary nature.


\section{Stellar astrophysical parameters: fitting PARSEC isochrones} \label{sec:benchmarking}



Using the PARSEC version 1.2S CMD web interface\footnote{http://stev.oapd.inaf.it/cmd}, we obtain families of isochrones in the Gaia DR3 photometric system. The range of age and abundance values is based on literature values from isochrone fitting to stars in the Hyades open cluster as compiled by \cite{Brandner2023}, i.e.\ log age [yr] = 8.70 to 8.95, and [M/H] = 0.10 to 0.25. We assume an average visual extinction of A$_{\rm V} = 3.1$\,mmag \citep{Taylor2006}. 
We use $\chi ^2$ minimization to identify the best fitting non-rotating PARSEC isochrones to the 80 brightest (G$_{\rm abs} \le 5.5$\,mag) bonafide single upper main sequence and main sequence turn-off stars, which yields $\log$ age [yr] = 8.89$\pm$0.01, and [M/H] = 0.18$\pm$0.03. 

\begin{figure}[ht!]
\plotone{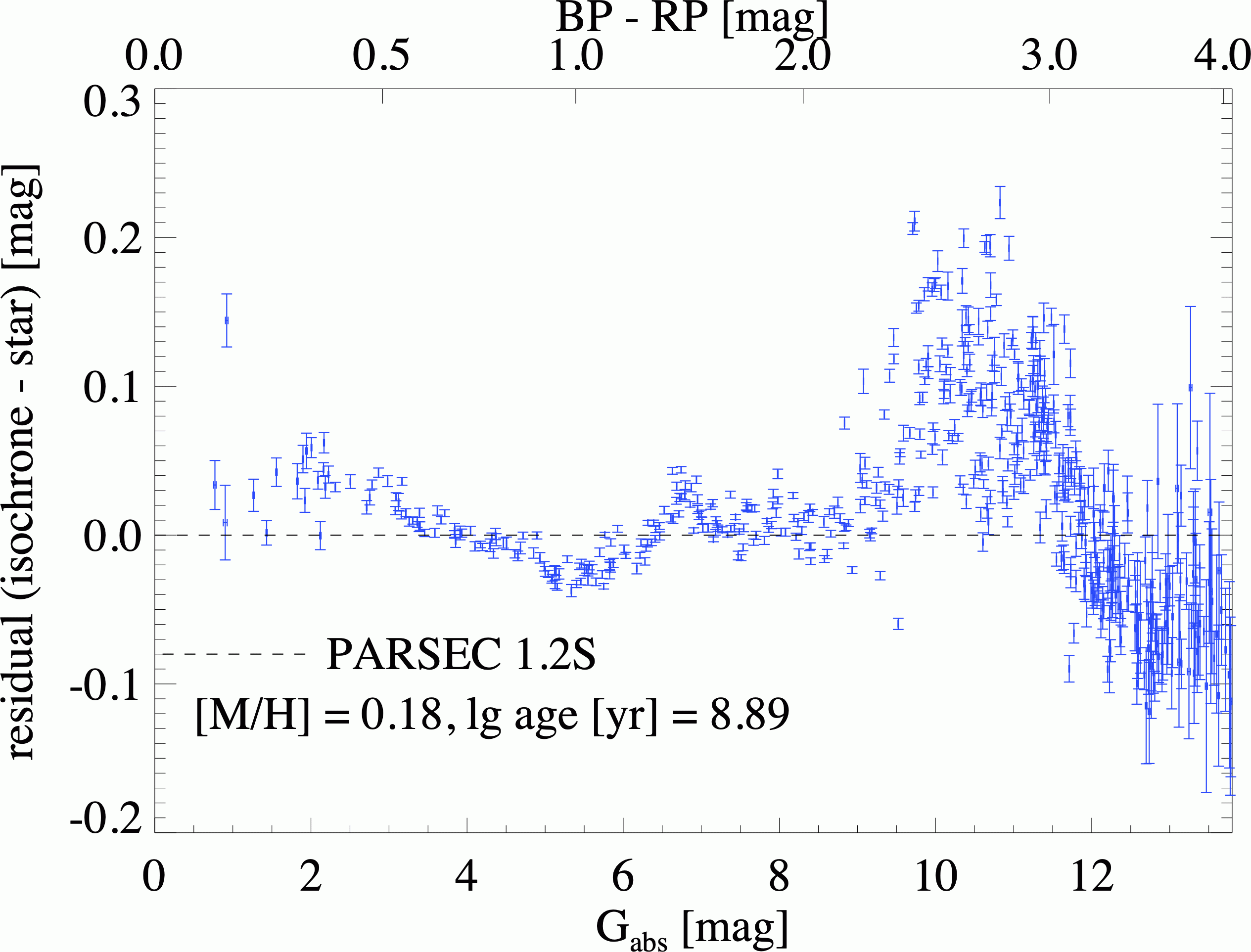}
\caption{Minimum distance in color-magnitude space of each single star from the best fitting isochrone, plotted against the absolute magnitude (lower abscissa) and color (upper abscissa) on the isochrone. A positive residual corresponds to stars, which are brighter (redder) than predicted by the isochrone. A negative residual corresponds to stars, which are fainter (bluer) than the isochrone. The uncertainties for each source correspond to the Gaia uncertainties in absolute G magnitude, and quadratically added uncertainties in absolute G magnitude and BP-RP color, respectively. 
\label{fig:IsoResiduals}}
\end{figure}

In Figure \ref{fig:IsoResiduals} we show the minimum distance in color-magnitude space of each star to the best fitting isochrone. Overall the residuals between the observed stellar sequence and the PARSEC isochrone are considerably smaller than is the case for the best fitting MESA 1.2 isochrone \citep{Brandner2023}. Noticeable is the systematic deviation for stars with 10\,mag$<$G$_{\rm abs}<$12\,mag. Stars in this brightness range also exhibit a large scatter, which is not explained by the intrinsic photometric uncertainty of the Gaia DR3 measurements.

\begin{figure*}[ht!]
\plotone{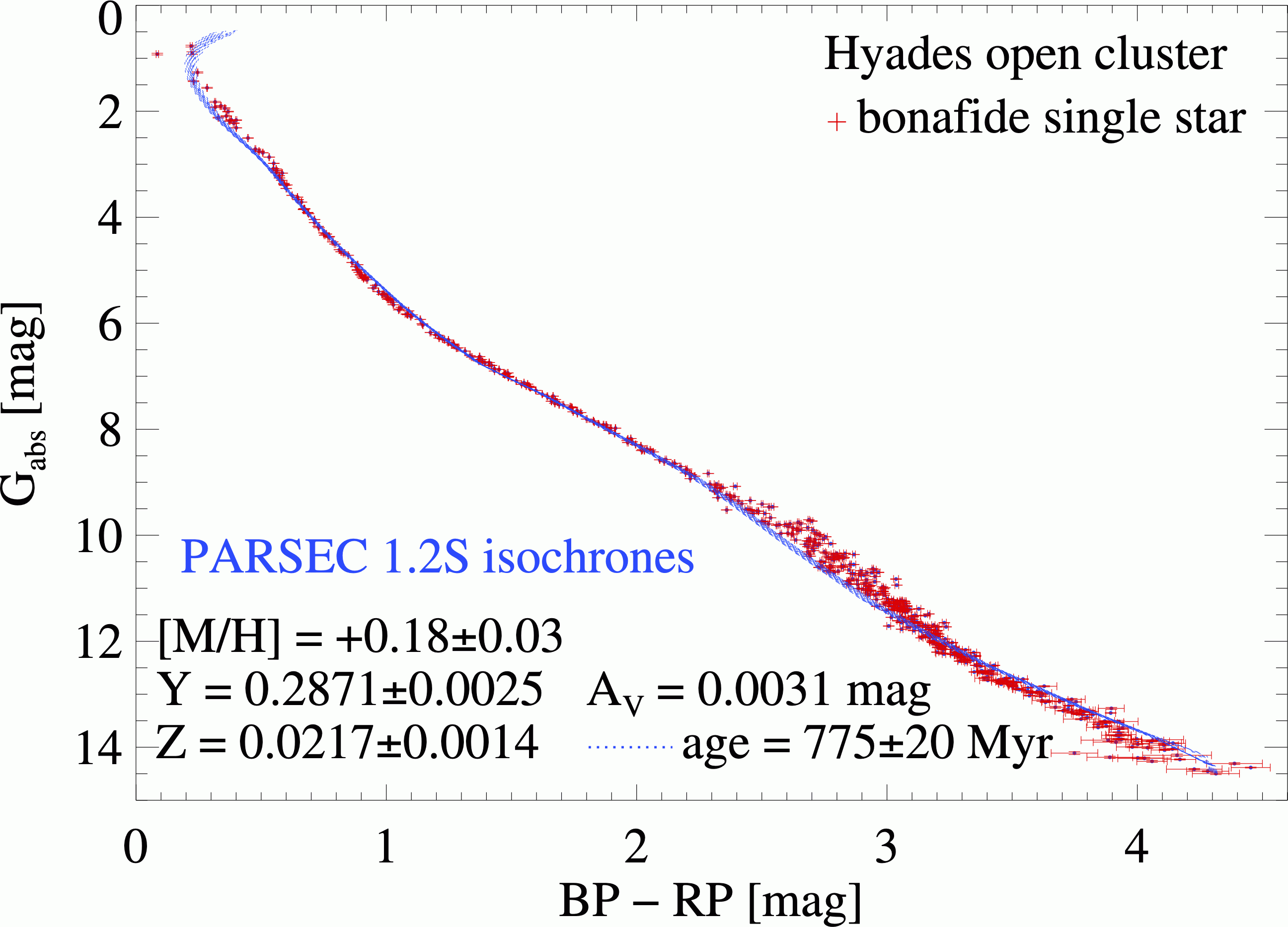}
\caption{Color-absolute magnitude diagram of $\approx$600 bonafide single star candidate members of the Hyades open cluster based on Gaia DR3. Overall, the family of PARSEC isochrones for [M/H]=$+0.18\pm0.03$, age = $775\pm 20$\,Myr, and A$_{\rm V} = 3.1$\,mmag provides a very good fit to the observed main-sequence. In the color range $2.4$\,mag$\le $BP-RP$\le$3.2\,mag ($\approx$0.22 to 0.40\,M$_\odot$) stars are systematically brighter (respectively redder) than predicted by the isochrones.
\label{fig:HyaPARSEC}}
\end{figure*}

Figure \ref{fig:HyaPARSEC} shows the single star main sequence of the Hyades with the family of best-fitting isochrones overplotted. We use the point of minimum distance on the isochrones to determine T$_{\rm eff}$, $\log$\,g, $\log$\,L, and (present-day) mass, and the associated uncertainties for each star, and present the results in Table \ref{tab:Hya_single_tab}.

\begin{deluxetable*}{rlllhhrhhhhhrrrrrrrr}
\tabletypesize{\scriptsize}
\tablewidth{0pt} 
\tablecaption{Astrophysical parameters of bonafide single stars in the Hyades open cluster \label{tab:Hya_single_tab}}
\tablehead{
\colhead{GAIA DR3 ID} & \colhead{RA}& \colhead{DEC} & \colhead{d} & \nocolhead{e_d} & \nocolhead{E\_d} & \colhead{G} & \nocolhead{e\_G} & \nocolhead{BP} & \nocolhead{e\_BP} & \nocolhead{RP} & \nocolhead{e\_RP} & \colhead{mass} & \colhead{$\sigma_{\rm mass}$}  & \colhead{log T$_{\rm eff}$} & \colhead{$\sigma_{\rm logTeff}$} & \colhead{log L} & \colhead{$\sigma_{\rm log L}$} & \colhead{log g} & \colhead{$\sigma_{\rm log g}$} \\
\colhead{} & \colhead{(deg)} & \colhead{(deg)} & \colhead{(pc)} & \nocolhead{(pc)} & \nocolhead{(pc)}& \colhead{(mag)} & \nocolhead{(mag)} & \nocolhead{(mag)}& \nocolhead{(mag)}& \nocolhead{(mag)}& \nocolhead{(mag)} & \colhead{(M$_\odot$)}  & \colhead{(M$_\odot$)} & \colhead{(K)}  & \colhead{(K)} & \colhead{(L$_\odot$)}  & \colhead{(L$_\odot$)} & \colhead{(cm/s$^2$}  & \colhead{(cm/s$^2$)}\\
} 
\startdata 
    395696646953688448&   0.653134&  51.945348&  59.611&  59.565&  59.655&  10.7791&  0.0001&  11.4353&  0.0004&  10.0031&  0.0003&  0.7431&  0.0025&  3.6508&  0.0004& -0.7832&  0.0021&  4.6464&  0.0008\\
    393017579491591168&   2.016317&  47.275403&  64.815&  64.423&  65.115&  17.0372&  0.0010&  19.2764&  0.0412&  15.6597&  0.0034&  0.1691&  0.0036&  3.4495&  0.0020& -2.6195&  0.0198&  5.0443&  0.0052\\
    420637590762193792&   2.509754&  55.443035&  83.610&  82.524&  84.646&  18.4767&  0.0017&  20.9101&  0.1079&  17.0266&  0.0074&  0.1460&  0.0054&  3.4278&  0.0063& -2.8203&  0.0554&  5.0717&  0.0098\\
    385502112574538624&   3.829797&  43.743120&  42.821&  42.780&  42.859&  14.3729&  0.0007&  16.1299&  0.0045&  13.1075&  0.0016&  0.3025&  0.0032&  3.4899&  0.0008& -2.1138&  0.0064&  4.9403&  0.0046\\
    394413482520591744&   4.297654&  49.937840&  68.621&  68.309&  68.863&  16.6537&  0.0008&  18.6973&  0.0163&  15.3238&  0.0020&  0.1960&  0.0033&  3.4621&  0.0012& -2.4837&  0.0113&  5.0242&  0.0048\\
    394913657235233664&   4.962093&  50.735753&  65.757&  65.377&  66.139&  17.0534&  0.0011&  19.2637&  0.0314&  15.6874&  0.0028&  0.1703&  0.0032&  3.4502&  0.0016& -2.6123&  0.0159&  5.0434&  0.0050\\
\enddata
\tablecomments{Table 1 is published in its entirety in the machine-readable format. The first six entries, and omitting some of the columns with uncertainties on the distance estimates, and part of the Gaia DR3 photometry, are shown here for guidance regarding its form and content.}
\end{deluxetable*}

\section{Discussion} \label{sec:discussion}

The inherently exquisite photometric and astrometric precision of Gaia DR3 reveals a very well defined and tight single star sequence in the Hyades. Families of PARSEC version 1.2S isochrones explain this as a stellar population with a small intrinsic spread in metallicity and age, and thus provide very accurate measurements of the absolute metallicity and age of the Hyades stellar open cluster in this frame of reference (Figure \ref{fig:HyaPARSEC}).

The strongest systematic deviations between observations and isochrones are in the mass range 0.22 to 0.40\,M$_\odot$, where stars are systematically brighter and/or redder than predicted by the isochrone by 0.02 to 0.2\,mag, and below 0.18\,M$_\odot$, where stars are fainter and/or bluer by 0.01 to 0.1\,mag. This is akin to a mismatch found between dynamical mass estimates and masses derived from isochrones for M-dwarf binaries and multiple system in the solar neighborhood, like, e.g., GJ\,2060, which is a member of the 50 to 100\,Myr old AB\,Dor moving group \citep{Rodet2018}, or 2MASS J10364483+1521394, which has an age of 400 to 600\,Myr \citep{Calissendorff2017,Calissendorff2018}.

For masses between 0.4 and 2\,M$_\odot$, the observed stellar sequence is in excellent agreement with the isochrones, and never deviates by more than 0.03\,mag in color-brightness space. This is different from the findings by \cite{Jaehnig2019}, who studied 69 presumed members of the Hyades open cluster in the mass range 0.5 to 1.0\,M$_\odot$. They report 15 stars with a radius inflation $\ge$10\%  with respect to a ``nominal'' isochrone. Seven of these 15 stars are in the GCNS sample of likely members of the Hyades. We find that two of these (2MASS J04084015+2333257, 2MASS J04115620+2338108) have problematic photometry in at least one of the Gaia photometric bands. According to \cite{Jaehnig2019} four of the remaining five stars are binary stars (2MASS J04175061+1828307, 2MASS J04181077+2317048, 2MASSJ04322565+1306476, 2MASSJ 04491296+2448103). Gaia DR3 detects evidence for orbital motion (i.e.\ RUWE$>$1.4) in 2MASS J04175061+1828307 and 2MASS J04491296+2448103. In \cite{Brandner2023} we flag 2MASS J04235070+0912193 as a probably binary stars due to its location on the binary sequence. \cite{Jaehnig2019} assign a radius inflation of $\Delta$R/R$=23.0\pm$6.6\% to this star. Observations at high angular resolution, possibly combined with radial velocity monitoring could help to decipher the true nature of this star, i.e.\ if it is an actual single star subject to radius inflation, or a close, thus far unresolved binary star.

While the PARSEC isochrones overall provide a considerable better fit to the Hyades single star sequence than, e.g., the MESA 1.2 isochrones, we note some caveats:

i) PARSEC version 1.2S does not consider stellar rotation. Stellar rotation should affect in particular the upper main sequence and main-sequence turn-off region in the Hyades, which is the region also most sensitive to the isochronal age.\footnote{\cite{Gossage2018} find that MESA models considering rotation result in lower effective temperature due to gravity darkening. When fitting MESA isochrones in the TYCHO photometric system to the Hyades, they derive a lower age estimate for the models with $\frac{\Omega}{\Omega_{\rm c}} = 0.3$ than for the non-rotating models.} For the majority of the stars in our sample, the effect on the determined astrophysical parameters should be minimal.
We note that stellar rotation is considered in the PARSEC evolutionary tracks and isochrones starting with version 2.0. However, at the time the present research was carried out, PARSEC version 2.0 was limited to $[M/H] \le +0.07$, which is below the canonical metal abundance estimates for the Hyades.

ii) the [He/H] abundance in the PARSEC models follows the solar-scaled Y=0.2485+1.78Z relation, while there is some evidence in the literature for a super-solar He abundance, see, e.g., the discussion in \cite{Tognelli2021}. 
A consequence of an underestimate of the He abundance would be an underestimate of the energy production for lower mass Hyades members, which still remember their primordial He abundance (i.e.\ these stars should not yet have reached perfect equilibrium on the p-p I $^3$He chain of nuclear fusion, see, e.g., \cite{Baraffe2018}).

iii) the astrometric and photometric identification of likely binary and multiple system is most efficient for approximately equal brightness and equal mass systems, i.e. systems with mass ratios $\ge$0.5 (see, e.g., \cite{Elson1998MNRAS}). More extreme brightness and mass ratio binary systems should still be present in our sample. 

iv) our sample is ignorant about stellar activity and variability, and individual extinction for each star. Here multi-wavelength SED fitting, as might be facilitated by the next generation of space-based infrared surveys like Euclid and Roman, should provide vastly improved characterizations.

Despite these limitations, the sample should be useful for a multitude of applications like benchmarking stellar evolutionary models, or the testing and verification of the calibration of large surveys such as the Gaia Astrophysical Parameters Inference System (APSIS, see, e.g., \cite{Fouesneau2022}). The table with astrophysical properties of $\approx$600 bonafide single stars in the Hyades  could also be useful to calibrate, e.g., different spectroscopic surveys with little or no internal overlap in their samples, against each other by using sources in common with our sample as a reference.
The sample also provides a homogeneous characterisation of the exoplanet hosts in the Hyades open cluster (see, e.g., \cite{Mann2016,Mann2018,Ciardi2018,Livingston2018,Vanderburg2018}), which facilitates a differential comparison of exoplanet properties in an open cluster.  

For stars in the mass range exhibiting radius anomalies, a more detailed study of rotation periods and activity levels is required. This might reveal insights into the stellar structure, and could probe the region susceptible to the convective kissing instability \citep{vanSaders2012,Baraffe2018,Mansfield2021,Mansfield2022}.






\begin{acknowledgments}
This work has made use of data from the European Space Agency (ESA) mission
{\it Gaia} (\url{https://www.cosmos.esa.int/gaia}), processed by the {\it Gaia}
Data Processing and Analysis Consortium (DPAC,
\url{https://www.cosmos.esa.int/web/gaia/dpac/consortium}). Funding for the DPAC
has been provided by national institutions, in particular the institutions
participating in the {\it Gaia} Multilateral Agreement.
\end{acknowledgments}

%

\vspace{5mm}
\facilities{Gaia}

\bibliography{lit}{}
\bibliographystyle{aasjournal}



\end{document}